\documentclass[twocolumn,showpacs,preprintnumbers,amsmath,amssymb,prl,aps]{revtex4-1}

\usepackage{graphicx}

\newcommand{\be}{\begin{equation}}
\newcommand{\ee}{\end{equation}}

\newcommand{\ket}[1]{\ensuremath{| #1 \rangle}}
\newcommand{\ovl}[2]{\ensuremath{\langle #1 | #2 \rangle}}

\newcommand{\ie}{{\it i.e.}}
\newcommand{\eg}{{\it e.g.}}

\newcommand{\hier}{\ensuremath{\nu}}
\renewcommand{\cup}{\ensuremath{v}}
\newcommand{\stime}{\ensuremath{\mathcal{T}}}

\begin{document}

\title{Efficient and coherent excitation transfer across disordered molecular networks}

\author{Torsten Scholak$^1$, Fernando de Melo$^{2,1}$, Thomas Wellens$^1$, Florian Mintert$^1$, and Andreas Buchleitner$^1$}

\date{\today}


\affiliation{%
  $^1$ Physikalisches Institut, Albert-Ludwigs-Universit\"at Freiburg, Hermann-Herder-Str.~3, D-79104 Freiburg, Germany\\%
  $^2$ Instituut voor Theoretische Fysica, Katholieke Universiteit Leuven, Celestijnenlaan 200D, B-3001 Heverlee, Belgium%
}

\begin{abstract}
We show that finite-size, disordered molecular networks can mediate
highly efficient, coherent excitation transfer
which is robust against ambient dephasing
and associated with strong multi-site entanglement.
Such optimal, random molecular conformations may explain
efficient energy transfer in the photosynthetic FMO complex.
\end{abstract}

\maketitle

Recently, a vivid debate arose on the physical mechanisms underlying efficient transport in organic molecules. In particular, the excitation transfer from the photoreceptor to the chemical reaction center in photosynthetic light harvesting complexes succeeds with astonishingly 
high transfer efficiency.
Since many of these biological systems exhibit disorder and are coupled to noisy environments, arguably all 
models \cite{Mohseni:2008zr,Caruso:2009vn} so far build on the fundamental hypothesis that disorder induces destructive interference
in the coherent quantum evolution. 
The latter, in turn, hinders transport \cite{Anderson:1958kx,Kramer:1993yb}, what can only be overcome by added noise, such as to restore 
the classically diffusive behavior. 
However, this hypothesis is valid only in the {\em thermodynamic limit}, i.e. for very large molecular structures, 
while coherent transport across 
{\em finite-size} disordered samples is 
characterized by large fluctuations under configurational variations \cite{Anderson:1958kx,Kramer:1993yb}. 
The photosynthetic FMO complex, as one of the most carefully studied examples \cite{Sarovar:2010uq}, is clearly 
a very {\em finite} molecular complex, {\em far from the thermodynamic limit}. New experimental data additionally
provide clear evidence that excitation transfer is predominantly coherent even at room 
temperature \cite{Engel:2007tg,Lee:2007hc,Panitchayangkoon:2010uq}, on {\em transient} time 
scales (of the order of a 
few $100 \textrm{fs}$) much shorter than the typical environment-induced decoherence times \cite{Cheng:2009ek}. 
Hence, the cause of the observed transport efficiency must be rooted in general properties of coherent 
quantum dynamics on {\em finite} molecular 
networks, on time scales {\em shorter} than those on which 
environmental decoherence fully develops its detrimental influence \cite{Logan:1987qf,Blumel:1991vn,Steck:2000ys}. As we will show,
rare incidences of {\em constructive rather than destructive} interference of transition amplitudes from the photoreceptor to 
the reaction center indeed do provide a possible explanation for these observations, and enable strictly coherent 
 transport efficiencies up to $100 \%$.

\begin{figure*}%
\centering
\includegraphics{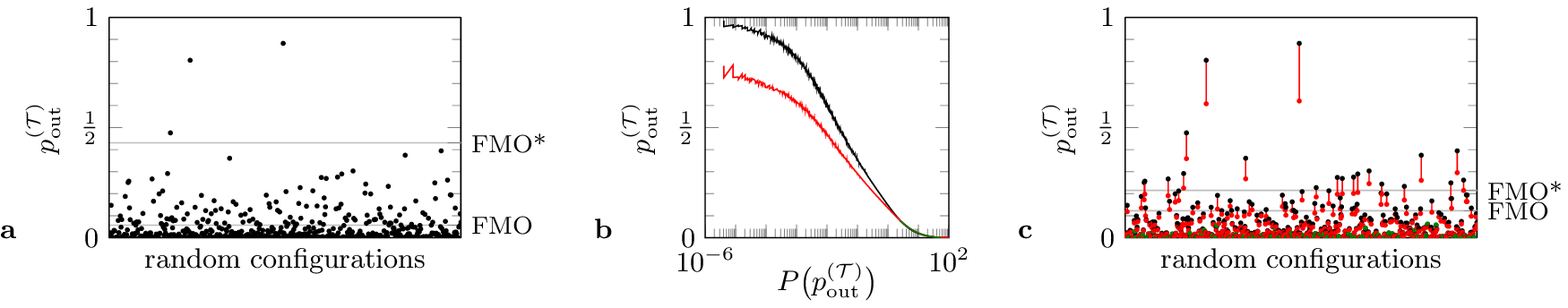}%
\caption{(Color online) {\bf a} Fluctuation of the transfer efficiency $p_\text{out}^{(\stime)}$
from input to output, for $500$ different random conformations of $N=7$ sites (dots).
Horizontal grey lines indicate the transfer efficiency of the experimentally infered \cite{[{Hamiltonian for P.~aestuarii, taken from: }] [{, tables 2 (MEAD) and 4 (right side, trimer).}]Adolphs:2006ve}
FMO Hamiltonian, as well as that of the optimal configuration FMO* 
compatible with the experimental error margin.
{\bf b} Probability densities $P\bigl(p_\text{out}^{(\stime)}\bigr)$ of the transfer efficiency $p_\text{out}^{(\stime)}$
for $2.5 \times 10^8$ different conformations.
For fully coherent dynamics (black curve)
the mean value of $p_\text{out}^{(\stime)}$ amounts to $4.9 \%$,
and only $4.5$ out of 
$10^6$ configurations provide efficiencies larger than $90 \%$.
Under local dephasing
(red and green curve) the mean efficiency drops to $3.9 \%$.
{\bf c} Gains (green) and losses (red) of the transfer efficiency with dephasing.
}
\label{fig:fluct}
\end{figure*}

\begin{figure}[b]%
\centering
\includegraphics{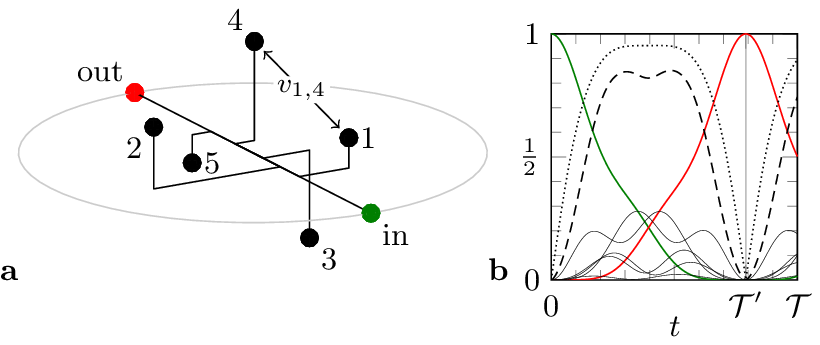}%
\caption{%
(Color online) {\bf a} Optimal spatial configuration of $N=7$ sites offering fast, robust,
and complete transport from input to output.
{\bf b} Time evolution of the on-site probabilites $|\ovl{i}{\psi(t)}|^2$
generated by the Hamiltonian defined by {\bf a}.
$i$ is either the input site \protect\includegraphics{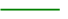},
the output node \protect\includegraphics{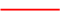},
or an intermediate site \protect\includegraphics{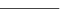}.
At time $\stime'$, only the output site is populated.
At intermediate times $t<\stime'$, the excitation is spread over several sites,
leading to high values of the bipartite \protect\includegraphics{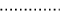}
and quadripartite \protect\includegraphics{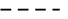}
entanglement, see text.
}
\label{fig:best}
\end{figure}

With some physical abstraction, a light harvesting complex can be viewed as a fully connected, disordered finite graph with
 $N$ vertices. Under the assumption of purely coherent transport, it is the spatial distribution of these which fully controls the 
 relative phases of the transition amplitudes which need to be added coherently to infer the 
 transport efficiency from the input to the output site. Optimal transport efficiency is then equivalent to 
 molecular 
 conformations with strictly constructive interference of all these amplitudes -- just conversely as in the case of disorder-induced
 destructive interference which dominates in the thermodynamic limit. On a finite molecular network, this represents an optimization 
 problem which can be solved by evolution, and suggests a statistical analysis, as follows. 

Coherent transport of a single excitation 
across a sample of molecular sites is generated by the Hamiltonian
\be
  H=\sum_{i \neq j = 1}^{N} \cup_{i,j}\ \sigma_+^{(j)}\sigma_-^{(i)}\, ,
\ee
where $\sigma_+^{(j)}$ and $\sigma_-^{(i)}$
mediate excitations and de-excitations of sites $j$ and $i$
from the local electronic ground state
to the local excited state and vice versa, respectively.
The excitation transfer $\sigma_+^{(j)}\sigma_-^{(i)}$ from site $i$ to site $j$ 
has a strength $\cup_{i,j}=\cup_{j,i}$ which depends on the specific nature of the inter-site coupling
-- that we assume to be of resonant (isotropic) dipole type,
$\cup_{i,j}=\alpha/r_{i,j}^{3}$, with $r_{i,j} = |\vec r_i-\vec r_j|$
and $\vec r_j$ the position vectors of individual sites.
Input and output site define the poles of a sphere of diameter $d$ which,
via the coupling constant $\cup_{\textrm{in},\textrm{out}} = \alpha/d^3$, 
sets the natural time-scale of the dynamics induced by $H$.
The positions of the remaining molecular sites are 
randomly (uniformly) chosen within this sphere, 
what induces a random distribution of the remaining $\cup_{i,j}$.

To assess the probability for complete and rapid transfer
of an excitation from the input to the output site
we sample over different spatial configurations.
Our figure of merit is the maximum probability
-- henceforth 
``transport efficiency'' --
\be
  p_\text{out}^{(\stime)} = \max_{t \in [0, \stime]} |\ovl{\text{out}}{\psi(t)}|^2
  \label{eq:transportefficiency}
\ee
that an excitation injected at input is received at output
after times no longer than 
$\stime = 0.1 \times \pi/(2|\cup_{\textrm{in},\textrm{out}}|)$.
This is one tenth of the time span the full excitation transfer would require
if no intermediate sites were present.
Note that this
specific choice of $\stime$ is immaterial for our subsequent discussion,
provided $\stime$ is sufficiently small as compared to
the time scale set by $\cup_{\textrm{in},\textrm{out}}$,
and long enough 
to allow
maximum values of $p_\text{out}^{(\stime)}$ close to unity.
This also implies (and is confirmed by inspection of the underlying numerical data) that large values of the transport efficiency as defined in \eqref{eq:transportefficiency} imply large values of the time-average of $|\ovl{\text{out}}{\psi(t)}|^2$ over $[0, \stime]$
(an efficiency quantifier used elsewhere \cite{Mohseni:2008zr}),
and vice versa. Only on time scales which are long with respect to $\stime$
could both quantifiers lead to different predictions. On such time scales, however, coherent effects will fade out.

Fig.~\ref{fig:fluct}{\bf a} shows the variation of 
the transport efficiency $p_\text{out}^{(\stime)}$
for a sample of $500$ different random distributions of
$N=7$ sites as in the FMO complex \cite{Cheng:2009ek}.
$p_\text{out}^{(\stime)}$ fluctuates wildly for different random configurations,
as anticipated in our motivation above.
Remarkably, very high transport efficiencies can be achieved as evident from Fig.~\ref{fig:fluct}{\bf b},
where the probability density of $p_\text{out}^{(\stime)}$
obtained for a sample of $2.5 \times 10^8$ realizations is shown in black.
Efficiencies above $90 \%$ are achieved for approx.\ five configurations out of a million,
despite the fact that the average efficiency is only $4.5 \%$.
Therefore, evolution can choose from truly exceptional,
albeit statistically unlikely molecular conformations 
with excellent transport properties.
Such a configuration is depicted in Fig.~\ref{fig:best} where
-- in contrast to the random choice of conformations in Fig.~\ref{fig:fluct} --
we have maximized $p_\text{out}^{(\stime)}$ by iterative optimization of the sites' positions.
Thereby, we find certain conformations which achieve even $100 \%$ transport efficiency.
The example shown in Fig.~\ref{fig:best}{\bf a} spells out that
optimal arrangements are
asymmetric and non-periodic, hence far from trivial (\eg, lattice-like, collinear) structures.  

To assess the efficiency of the actual FMO complex,
in comparison to our present results,
we employ an approximate Hamiltonian
\cite{[{Hamiltonian for P.~aestuarii, taken from: }] [{, tables 2 (MEAD) and 4 (right side, trimer).}]Adolphs:2006ve}
inferred from experimental data,
and obtain a transport efficiency of only $5.7 \%$ in a time window of $1.6 \times 10^{-13} \mathrm{s}$,
hence close to the average value of our random model.
The time window was here defined by the coupling strength between chromophores $1$ and $3$,
$\stime = 0.1 \times \pi/(2|\cup_{1,3}|)$.
However, variation 
of the off-diagonal (diagonal) matrix elements
by at most $3.2 \times 10^{11} h \mathrm{s}^{-1}$
($17 \times 10^{11} h \mathrm{s}^{-1}$)
-- what is the absolute error margin deduced from the experimental data
\cite{[{Hamiltonian for P.~aestuarii, taken from: }] [{, tables 2 (MEAD) and 4 (right side, trimer).}]Adolphs:2006ve} -- 
is compatible with an alternative,
optimal configuration (FMO*) with a transport efficiency
of $43.1 \%$, see Fig.~\ref{fig:fluct}{\bf a}
($h$ is Planck's constant)!
Furthermore, much as the optimal configuration depicted in Fig.~\ref{fig:best}{\bf a},
this configuration's efficiency is robust
under statistical variations with a spread
of $10^{11} h \mathrm{s}^{-1}$
($5.4 \times 10^{11} h \mathrm{s}^{-1}$)
on the FMO* Hamiltonian's off-diagonal (diagonal) elements,
in the sense that such variation yields
a Gaussian distribution $43.1 \pm 5 \%$
of the transport efficiency around the optimum.

Let us now consider the same transport problem in the presence of environmental noise. 
Fig.~\ref{fig:fluct}{\bf c} shows the efficiencies $p_\text{out}^{(\stime)}$
for the same statistical sample as in Fig.~\ref{fig:fluct} {\bf a},
under local dephasing with a strong rate $\gamma=2/\stime$.
Cases in which $p_\text{out}^{(\stime)}$ is decreased by dephasing are highlighted in red,
enhancements of transport efficiency is depicted in green.
The plot very neatly spells out a clearly dichotomous impact of the environment:

  (i) Whenever constructive quantum interference enhances transport in the absence of environment coupling,
    the noise reduces the transport efficiency $p_\text{out}^{(\stime)}$ very considerably.

  (ii) In contrast, if quantum coherence suppresses transport in the strictly coherent case,
    dephasing will enhance $p_\text{out}^{(\stime)}$,
    though only very marginally so.

  (iii) Notwithstanding, even in the presence of the rather strong dephasing chosen for our simulation,
    those rare molecular conformations which provide efficient excitation transfer
    maintain this characteristic property under environmental coupling,
    just at reduced levels,
    and are still clearly distinct from those conformations which hinder transport.
    
The crossing of the probability densities obtained with and without noise,
see Fig.~\ref{fig:fluct}{\bf b},
identifies a level of $7.6 \%$ as the demarcation line between transport efficiencies
which are strongly reduced ($p_\text{out}^{(\stime)} > 7.6\%$)  or marginally enhanced ($p_\text{out}^{(\stime)} < 7.6\%$) by 
added noise.
This is again in qualitative accord with the reported data on the FMO Hamiltonian, as well as with its efficient variant FMO* 
introduced above: while the FMO efficiency increases from $5.7$ to $12.3$ percent, the FMO* efficiency is reduced from $43.1$ 
to $21.5 \%$, in the presence of noise.

Note that such dichotomous behavior
as identified here for {\em finite} systems on {\em transient} time scales
is {\em absent} in the thermodynamic limit of {\em infinite} systems and/or sufficiently long transport times,
where noise completely destroys quantum coherences
and tends to induce near-classical, diffusion-like behavior.
It is well-known that noise then enhances transport by suppressing
{\em destructive} quantum interference \cite{Logan:1987qf,Blumel:1991vn,Steck:2000ys,Mohseni:2008zr},
though can in general not compete with
the transport efficiencies brought about by constructive quantum interference
-- even when the environment coupling strengths are optimized \cite{Caruso:2009vn}.

\begin{figure*}%
\centering
\includegraphics{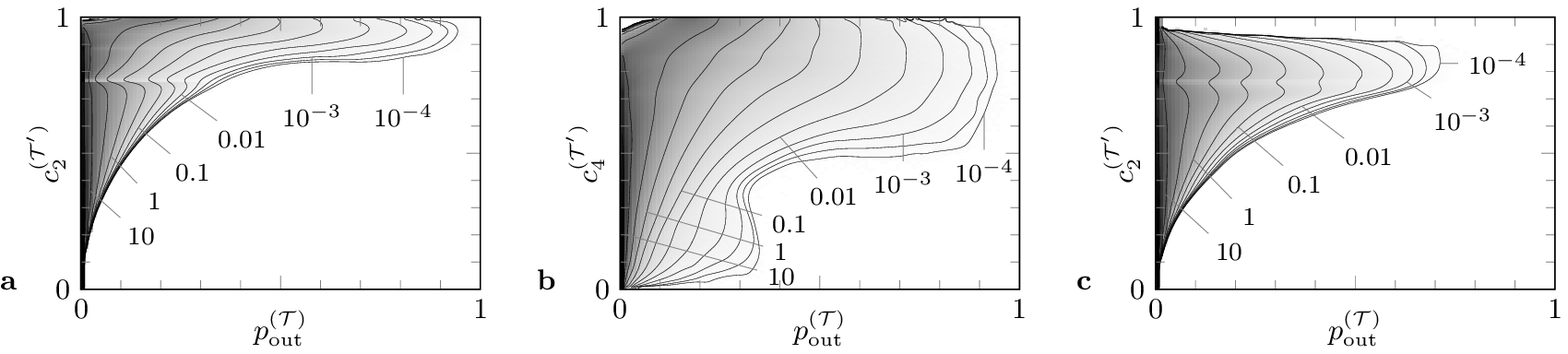}%
\caption{%
{\bf a} Contour plot of the conditional probability density $P\bigl(p_\text{out}^{(\stime)}\bigm|c_2^{(\stime')}\bigr)$, 
\ie\ the probability distribution of the transport efficiency $p_\text{out}^{(\stime)}$ across $7$ sites, 
given the generation of a certain maximal level $c_2^{(\stime')}$ of at least bipartite entanglement,
during the transfer time $\stime'$.
{\bf b} Same as {\bf a}, but for quadripartite entanglement $c_4^{(\stime')}$.
{\bf c} Same conditional probability distribution as in {\bf a},
when all sites are locally coupled to a dephasing environment,
with dephasing rate $\gamma = 2/\stime$.
In all three cases, large transport efficiencies require a minimum amount of entanglement.
}
\label{fig:contour}
\end{figure*}

Efficient quantum transport as observed above relies on the constructive interference of 
a large number of (input to output) transition amplitudes, reminiscent of efficient quantum algorithms.
Therefore, let us now address the question whether multi-site entanglement is of similar relevance for
the molecular transport problem as it is for 
quantum computation -- an issue of much 
recent interest
\cite{Thorwart:2009ve,Asadian:2010fk,Sarovar:2010uq}. 
In our present problem,
precisely one excitation propagates from input to output,
so that the transporting states are close relatives of the W-states \cite{Dur:2000vn}
-- a well-known class of entangled states of multipartite systems,
that are known to be particularly robust against decoherence induced by dephasing or spontaneous 
decay \cite{Carvalho:2004ij}. These states also provide a clear relation between the excitation's localization
and the system's entanglement properties: The latter can be characterized in terms of a hierarchy of 
quantities $\{c_\hier(\psi)\}_{\hier=2,\hdots,N}$ which are strictly positive
if entanglement in $\ket{\psi}$ is shared by at least $\hier$ sites, and vanish otherwise. In particular,
each of the $c_\hier(\psi)$ is a function of the statistical moments $M_k(\psi)=\sum_{j=1}^N|\langle j|\psi\rangle |^{2k}$, 
where $\{\ket{j}\}_{j=1,\ldots,N}$ is the canonical site basis.
In terms of the second moment $M_2(\psi)$
-- which is nothing but the inverse participation ratio \cite{Kramer:1993yb}
frequently used in statistical descriptions of complex quantum systems -- 
$c_2(\psi)$
(which is a multi-partite generalization \cite{Mintert:2005ys} of a standard bipartite entanglement measure \cite{Bennett:1996zr})
reads
\begin{subequations}
\be
  c_2(\psi)=\sqrt{\tfrac{1}{1 - 1/N}(1 - M_2(\psi))}\ .
\ee
Higher order $c_\hier(\psi)$ are analogous functions of the moments $M_1(\psi) = 1$ to $M_\hier(\psi)$.
For instance, the quadripartite measure reads
\be
  c_4(\psi) = \biggl(\frac{1 - 6 M_2 + 8 M_3 + 3 M_2^2 - 6 M_4}{1 - 6/N + 11/N^2 - 6/N^3}\biggr)^{1/4}\ .
\ee
\end{subequations}
Under purely coherent dynamics, multi-partite entanglement thus encodes the detailed localization properties 
of the excitonic wave function. 

With these tools at hand,
we can now correlate the transport efficiency with the multi-site entanglement
which is generated during the transport process.
Figs.~\ref{fig:contour}{\bf a} and \ref{fig:contour}{\bf b} show the probability density of the transport efficiency,
$P\bigl(p_\text{out}^{(\stime)}\bigm|c_2^{(\stime')}\bigr)$ and $P\bigl(p_\text{out}^{(\stime)}\bigm|c_4^{(\stime')}\bigr)$,
conditioned on the maximal bi- and quadripartite entanglement, $c_2^{(\stime')}$ and $c_4^{(\stime')}$,
which is generated during the exciton propagation from input to output:
\be
  c_\nu^{(\stime')} = \max_{t \in [0, \stime']} c_\nu(\psi(t))\ .
\ee
$\stime' \le \stime$ is the time at which the maximum output probability $p_\text{out}^{(\stime)}$ is reached,
see Fig.~\ref{fig:best}{\bf b}.
Clearly, {\em efficient transport necessarily requires strong entanglement}.
This is most prominently spelled out in Fig.~\ref{fig:contour}{\bf a},
where high transport efficiency (\eg\ $p_\text{out}^{(\stime)}>0.5$) is only reached
at high values of the entanglement ($c_2^{(\stime')}>0.8$) shared between at least two sites. 
Note the kink in the distribution visible at $c_2^{(\stime')} = \sqrt{7/12} \simeq 0.76$,
which corresponds to maximal entanglement between exactly two of $N=7$ sites.
Here, transport is inhibited since the excitation may be trapped in a singlet state between two sites
which are accidentally placed very close to each other.  

The correlation between transport and entanglement visible in Fig.~\ref{fig:contour}{\bf a}
prevails for higher orders of the $c_\hier^{(\stime')}$,
but is less pronounced for increasing $\hier$,
as evident from the exemplary case of $c_4^{(\stime')}$ in Fig.~\ref{fig:contour}{\bf b}. 
Here, although moderate transport efficiencies like $p_\text{out}^{(\stime)}\simeq 0.2$
are possible at very small values of $c_4^{(\stime')}$,
higher transport efficiency still requires a certain amount
of entanglement (i.e. $c_4^{(\stime')}>0.5$) between at least four sites. 

Finally,
to gauge the robustness of the observed correlation under decoherence,
Fig.~\ref{fig:contour}{\bf c} shows the conditional probability density
$P\bigl(p_\text{out}^{(\stime)}\bigm|c_2^{(\stime')}\bigr)$ 
estimated for mixed states \cite{Mintert:2005qf}
when the individual molecular sites are locally coupled to dephasing environments,
with the same decay rate $\gamma=2/\stime$ as in Fig.~\ref{fig:fluct}{\bf c}.
Fully consistent with our discussion of Fig.~\ref{fig:fluct}, 
the correlation between entanglement and transport efficiency remains qualitatively unaffected,
however with smaller transport efficiencies and entanglement levels
than in the strictly coherent case. This correlation is particularly remarkable here, since the above equivalence of
multi-site coherence and multi-site entanglement cannot be established any more 
under open system dynamics.

In conclusion, we have seen that very 
well defined molecular configurations, which can be found by iterative optimization, mediate highly efficient and robust transport across molecular networks
alike the FMO energy harvesting complex. Even in the presence of rather strong dephasing
does efficient excitation transfer due to constructive quantum interference remain a distinctive feature of these conformations. Efficient transport is furthermore conditioned on the build-up of strong inter-site entanglement in the course of the exciton transfer. 
This is clear evidence of the functional role of multi-site entanglement
on the level of biomolecular (quantum) dynamics.
Whether, beyond that, biology has ways to harvest
the statistical, non-local quantum correlations
between single excitation events at different sites of $W$-like states
on FMO-like functional units 
remains an intriguing question for future research.

We enjoyed illuminating discussions with Markus Tiersch and Simeon Sauer,
and acknowledge financial support by Alexander von Humboldt Foundation
and the Belgian Interuniversity Attraction Poles Programme P6/02 (F.d.M),
as well as by DFG (F.M.).

\end{document}